\documentclass[iop]{emulateapj}

\usepackage{comment}
\usepackage{enumerate}
\usepackage[T1]{fontenc}
\usepackage{pslatex}
\usepackage{epstopdf}
\usepackage{amsmath}
\usepackage{graphicx}
\usepackage{float}

\slugcomment{Accepted for publication in the Astronomical Journal}

\def\deg{^\circ }
\def\solar{_\odot }
\def\Sec{''}

\shorttitle{Strong Ram Pressure Stripping in NGC 4921}
\shortauthors{Kenney et al.}

\begin{document}

\title{
HST and HI Imaging of Strong Ram Pressure Stripping in the Coma Spiral NGC 4921:
Dense Cloud Decoupling and Evidence for Magnetic Binding in the ISM}
\author {Jeffrey D. P. Kenney\altaffilmark{1},
Anne Abramson\altaffilmark{1}, \&
Hector Bravo-Alfaro\altaffilmark{2,3}
}
%
\altaffiltext{1}{\scriptsize\
Yale University Astronomy Department,
P.O. Box 208101, New Haven, CT 06520-8101 USA jeff.kenney@yale.edu}

\altaffiltext{2}{\scriptsize\
Institut d'Astrophysique de Paris; CNRS/UPMC, 98bis, Boulevard Arago 75014, Paris, France}

\altaffiltext{3}{\scriptsize\
Departamento de Astronomia, Universidad de Guanajuato, Apdo. Postal 144, Guanajuato 36000, Mexico}


%
\begin{abstract}
Remarkable dust extinction features in the deep HST V and I images of the face-on Coma cluster spiral galaxy NGC 4921 show in unprecedented ways how ram pressure strips the ISM from the disk of a spiral galaxy.  
New VLA HI maps show a truncated and highly asymmetric HI disk with a compressed HI distribution in the NW, providing evidence for ram pressure acting from the NW.
Where the HI distribution is truncated in the NW region, HST images show a well-defined, continuous front of dust that extends over 90 degrees and 20 kpc. This dust front separates the dusty from dust-free regions of the galaxy, and we interpret it as galaxy ISM swept up near the leading side of the ICM-ISM interaction. We identify and characterize 100 pc-1 kpc scale substructure within this dust front caused by ram pressure, including head-tail filaments, C-shaped filaments, and long smooth dust fronts.  The morphology of these features strongly suggests that dense gas clouds partially decouple from surrounding lower density gas during stripping, but decoupling is inhibited, possibly by magnetic fields which link and bind distant parts of the ISM. 
\end{abstract}


\section {Introduction}

Ram pressure stripping (rps) is an important mechanism for removing gas, quenching star formation and driving galaxy evolution in sufficiently dense environments \citep{gg72, hg86, cay90, bravo00, kk04b, bg06, pog09, voll13a, cen14, bos14}.  
In recent years many large galaxies in both nearby and distant clusters have been found to have prominent one-sided gas tails and other clear evidence of active rps \citep{kvgv04,ovg05,chung07, yagi10, smith10, sun10, owers12, ebel14, jac14}. Many cluster spiral galaxies have
truncated gas and star formation distributions inside relatively undisturbed stellar disks \citep{cay90, bravo00, kk04b, chung09}, a signature of ongoing or past rps.
In high mass clusters (M$\sim$10$^{15}$ M$_{\odot}$) like Coma, spirals can be completely stripped 
\citep{bravo00,yagi10, smith10}. In medium mass clusters (M$\sim$10$^{14}$ M$_{\odot}$) like Virgo, spirals are generally partly stripped \citep{cay90, kk04b, chung09}, but dwarf galaxies get completely stripped, leading to dI-dE transformation \citep{bos08,kb12,ken14}. RPS is likely responsible for the lack of HI gas \citep{gp09,spek14} and star formation \citep{ein74,hain07,geha12} in satellite dwarf galaxies that are sufficiently close to their more massive host galaxy, although it has been questioned whether ram pressure alone in the Local Group is strong enough to do the job \citep{nbh11}. While rps is clearly important in clusters, we still wish to understand how far out in clusters it happens, and how much it is enhanced by shocks and substructure in the ICM \citep{kvgv04,tb08,owers12}. We still need to understand how important rps is in groups, either from the disk or the halo. Ram pressure stripping of halo gas, which is one form of starvation or strangulation \citep{larson80,bal00} is probably an important quenching mechanism although its impact remains unclear, in part, because we do not understand the physics of rps sufficiently well. 

In order to understand the impact of rps as SF quenching mechanism throughout the universe, we need to better understand the efficiency of stripping, e.g., relative to the simple  \citet{gg72} criterion, in which gas is stripped if the ram pressure $\rho$v$^2$ exceeds the gravitational restoring force. Several things may affect the stripping efficiency, including the duration of ram pressure 
\citep{voll01,rb07,jac07}, the disk-wind angle \citep{voll01,rb06,jac09}, the strength of hydrodynamical instabilities \citep{rbh06,rh08}, ISM substructure \citep{tb09,tb10}, and magnetic fields \citep{rus14,ts14}.
Of particular interest is the fate of dense star-forming gas clouds, which are generally too dense to directly strip, but are important for understanding the quenching and triggering of star formation during ram pressure. There is evidence that dense clouds can decouple from surrounding low density gas during stripping \citep{crowl05, voll05, voll08a, ak14} but we still do not know how easily this happens, under what physical conditions, how long clouds survive after decoupling, and how star formation is affected. Our understanding of how stripping actually happens on the ISM scales critical for star formation of 1-100 pc has been limited by a lack of high resolution and high sensitivity ISM observations in galaxies experiencing strong ram pressure. 

New insights into rps come from HST images of the Coma galaxy NGC~4921, which is a spectacular example of a massive (M$_{\rm B}$=-22) nearly face-on (i<30$\deg$) spiral galaxy being actively stripped by ram pressure.  NGC 4921 is located fairly close (25$'$=700 kpc=0.35R$_{vir}$) to the center of the Coma cluster. Star formation is weak so the galaxy is anemic.  In the Virgo cluster (16 Mpc, 1$''$=77pc) the best stripping cases are highly inclined \citep{kvgv04,crowl05,ovg05,abram11}, and there are no clear examples known of spirals being strongly stripped when they are viewed relatively face-on.  The Coma cluster (100 Mpc, 1$''$=462pc) is more massive than Virgo by an order of magnitude and has much stronger ram pressure, by virtue of both higher orbital velocities and higher ICM gas densities. While many galaxies in Coma  \citep{yagi10,smith10,fos12,yosh12} and other clusters  \citep{chung07,cor07,sun10,ken14,ebel14} exhibit spectacular tails of gas and young stars formed by rps, in order to understand the efficiency of stripping one needs to study the behavior of the ISM {\it in the disk} as it is being stripped. In face-on galaxies like NGC~4921 it is possible to see how the ISM in the disk is affected by ram pressure.

Inspection of the HST images of NGC~4921 reveals numerous dust extinction features 
with a type of substructure not previously seen or recognized in any other galaxy. They provide significant new insight into how the real 
multiphase substructured ISM in spiral galaxies responds to strong ram pressure. The HST images are deep ACS/WFC images in F606W (V, 62 ksec) and F814W (I, 37 ksec) taken from a Cepheid program (proposal ID 10842, PI Cook). A paper by \citet{tg11} mentioned but did not investigate the substructure of the dust features, as their work focused on stellar populations. A paper by \citet{carl13} discussed many dust features in the galaxy, and proposed that protruding dust filaments are magnetic ``outgrowths'' of spiral arms. Our interpretations of the dust substructure are very different from \citet{carl13}, as they do not consider the effects of ram pressure, which we believe is the primary cause of the interesting dust filaments. No one has yet studied the dust morphology of NGC 4921 to learn about the effects of ram pressure. 

The HST images have resolution of 0.05$''$ which corresponds to 30 pc at Coma. 
In this paper we show the image produced by Roberto Colombari (http://www.astrobin.com/users/rob77/), but rotated so that north is up. This image nicely highlights the contrast between the background starlight and the dust features we are interested in studying. Such high resolution images of the ISM will be a valuable guide to future simulations that can teach us how stripping acts in detail, how efficient gas stripping is, and ultimately how important gas stripping is in driving galaxy evolution in different environments.

\section{HI distribution and kinematics}

\begin{figure}[ht]
\epsscale{1.15}
\plotone{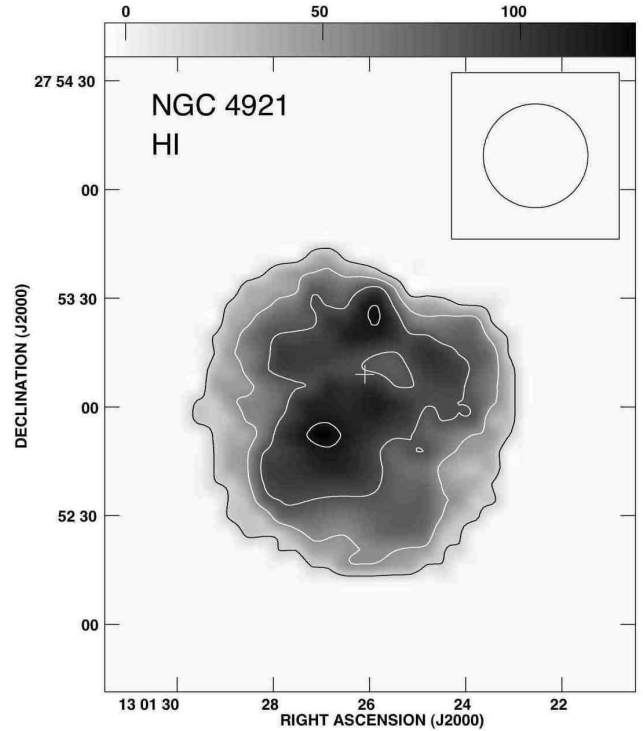}
\caption{
VLA HI intensity (moment 0) map of NGC 4921. 
Beam size (28.8$\Sec$x28.6$\Sec$) shown in upper right. Optical center marked with cross.
Contour levels are 1,3,5,7 times 17.5 Jy/beam m/s, or 2.6$\times$10$^{19}$ atom cm$^{-2}$.
Note highly asymmetric HI distribution with shortest HI extent and compressed HI contours in the NW.
\label{fig1}
}
\end{figure}

\begin{figure*}[ht]
\plotone{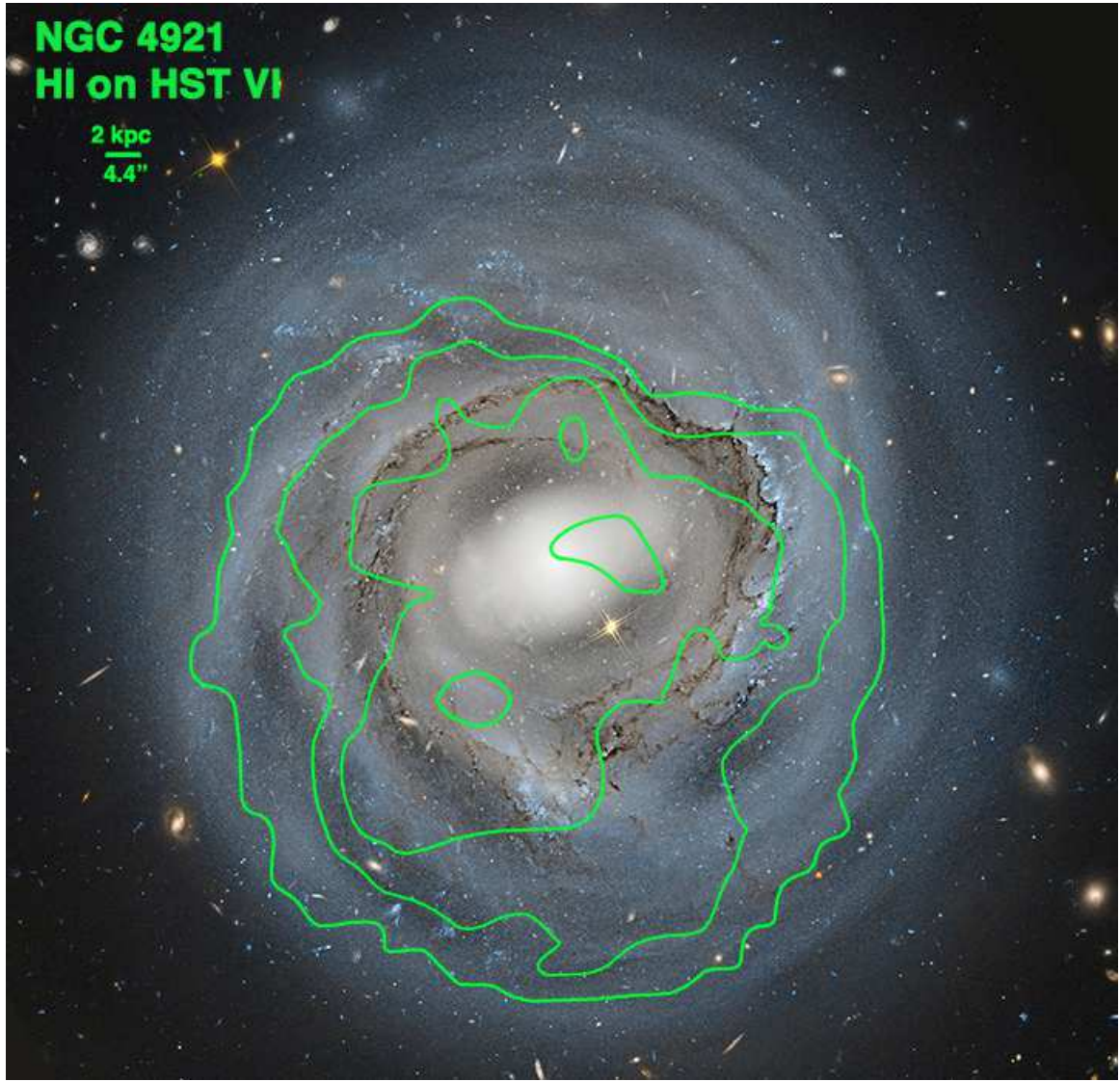}
\caption{
HI intensity contour map on color HST image (F606W+F814W) of NGC~4921. Contours as in Fig 1.
Note highly asymmetric HI distribution with shortest HI extent and compressed HI contours in the NW near the prominent dust lane.
\label{fig2}
}
\end{figure*}

\begin{figure}[ht]
\epsscale{1.2}
\plotone{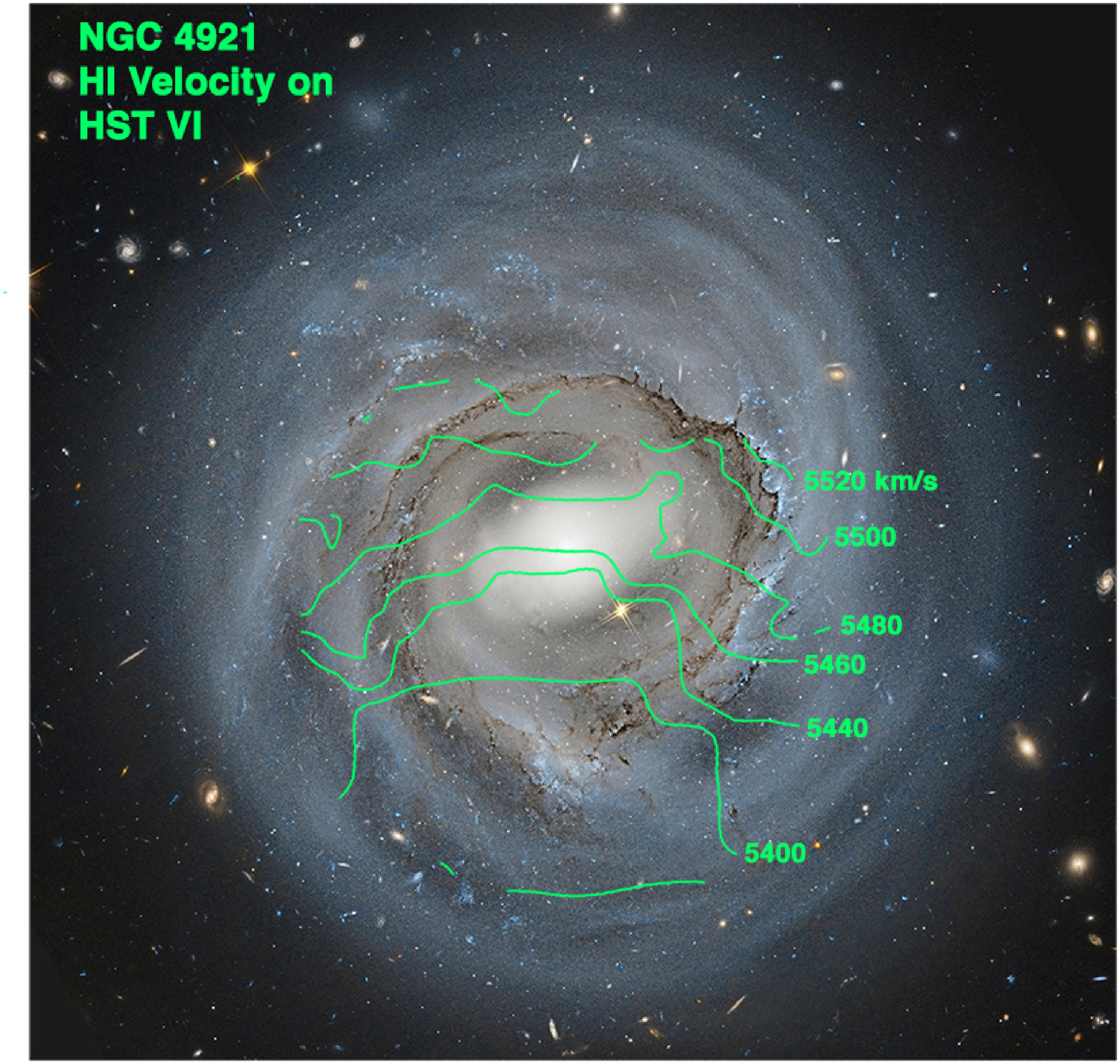}
\caption{
HI velocity contour (moment 1) map on color HST image (F606W+F814W) of NGC~4921. 
Note the outer isovelocity contours in the northern half of the galaxy curve towards the minor axis, indicating non-circular motions.
\label{fig3}
}
\end{figure}

We first present evidence from HI data that the ISM of NGC 4921 is being ram pressure stripped. HI was observed by the VLA in C and D arrays in 1994-99 and reported by \citet{bravo00,bravo01}, who detected a truncated asymmetric HI disk with more HI to the SE. 
We have reprocessed the data in order to improve the signal-to-noise
ratio and detection limit. We found that continuum subtraction
was a critical step in the data reduction. The regular procedure
goes through the search of line-free channels and then use AIPS
tasks (e.g. UVLIN) to subtract the continuum in the UV-plane.
In the present work we explored different approaches to define
our line-free channels and carried out the continuum subtraction
in the map-plane, using tasks SQASH and COMB. We found that
this strategy enables us to present maps of higher sensitivity
and resolution than those presented previously. 
In this paper we show maps from the combined C+D array data cube with a beam size of 28.8$\Sec$x28.6$\Sec$, made with nearly pure natural weighting (robust parameter of +5).  The rms in this new cube is 0.17 mJy/beam, which is improved by 15\% compared with the rms (0.20 mJy/beam) reported by \citet{bravo00}, with a larger beam of 39.8$\Sec$$\times$34.5$\Sec$. The new maps show a slightly more extended HI distribution than the older maps but confirm the deeply shrunken gas disk seen in \citet{bravo01}. The re-estimated HI flux is 0.73$\pm$ 0.02 Jy km/s, 20\% higher than the earlier value of 0.61$\pm$0.03 Jy km/s \citep{bravo00}. This gives a global HI deficiency of 1.0, implying that the galaxy has lost $\sim$90\% of its HI. 

Figure 1 shows the HI total intensity map, from the 0th moment of the data cube, and Figure 2 shows this map overlaid on the HST V band (606W) optical image. The HI distribution in NGC 4921 is truncated well within the optical disk and is highly asymmetric, extending to  53$\Sec$ (24 kpc, 0.77R$_{25}$) in the SE but only 28$\Sec$ (13 kpc , 0.40R$_{25}$) in the NW, a difference of nearly a factor of two. HI is least extended at PA = 335$\pm$5$\deg$ and most extended on the opposite side at 165$\pm$5$\deg$. This is the type of distribution expected from external pressure from PA = 335$\deg$. While there can be offsets between the wind and tail directions \citep{rbh06}, the fact that the tail direction is directly opposite the side with shortest HI extent and compressed HI contours suggests that the wind direction is not too different from PA=335$\deg$, which is roughly the direction toward the cluster center. 

Figures 1 and 2 show that a central deficit of HI extends from the center toward the NW side of the bar out to $\sim$12$''$ (5 kpc). Many spiral galaxies have HI deficits near the center or in the barred region \citep{ond89, walter08}. The SE side of the bar has a bit more HI than the NW side, and the maximum surface density in the galaxy is located just outside bar-dominated region in the SE, which is the downstream direction.  

Figure 3 shows the HI velocity map, from the 1st moment of the data cube, overlain on the HST V band (606W) optical image. The HI velocity field shows a distorted spider pattern. Nearly all the isovelocity contours curve downwards, as shown in the original \citep{bravo01} map. In the S half of the galaxy,  the isovelocity contours curve away from the minor axis, consistent with dominant rotational motions, and not highly disturbed. But in the N half of the galaxy,  the isovelocity contours curve {\it toward} the minor axis,  which is inconsistent with pure rotation, and implies that the outer gas in the N has strong non-circular motions, presumably due to ram pressure.

A comparison of the HST image with the HI map in Figure 2 reveals that the radial extent of HI is similar to that of the dust extinction. In viewing the HI-HST overlays one must keep in mind the much lower resolution of the HI map (28$''$) compared that that of HST (0.05$''$).

\section{Morphology of Leading Side Dust Front on Large Scales}

\begin{figure*}[ht]
\plotone{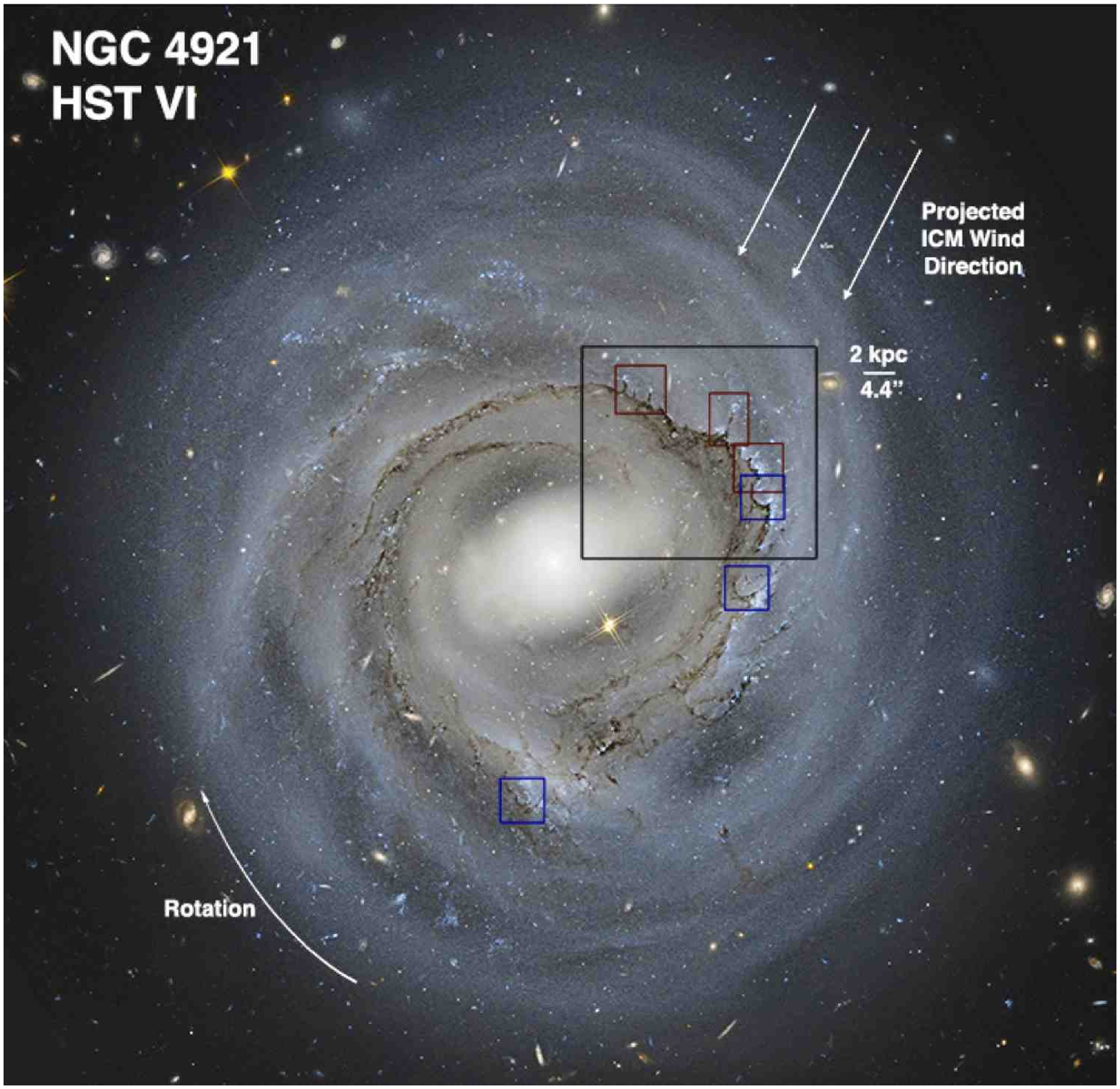}
\caption{
Deep HST V+I image of Coma spiral NGC 4921 from Cook Cepheid program (ID 10842), processed by Roberto Colombari. Pseudo 3-color image, with blue = F606W=V, red = F814W=I, green = (red+blue)/2.  The image highlights the dust extinction features, although appears bluer than the real galaxy. Arrows indicate the direction of galaxy rotation and the likely ICM wind angle. Boxes outline regions highlighted in other figures. The leading (NW) side of the disk shows the effects of strong ram pressure on the dust. The trailing side shows weaker effects from ram pressure, providing evidence for shielding effects from intervening ISM.
\label{fig4}
}
\end{figure*}

\begin{figure*}[ht]
\plotone{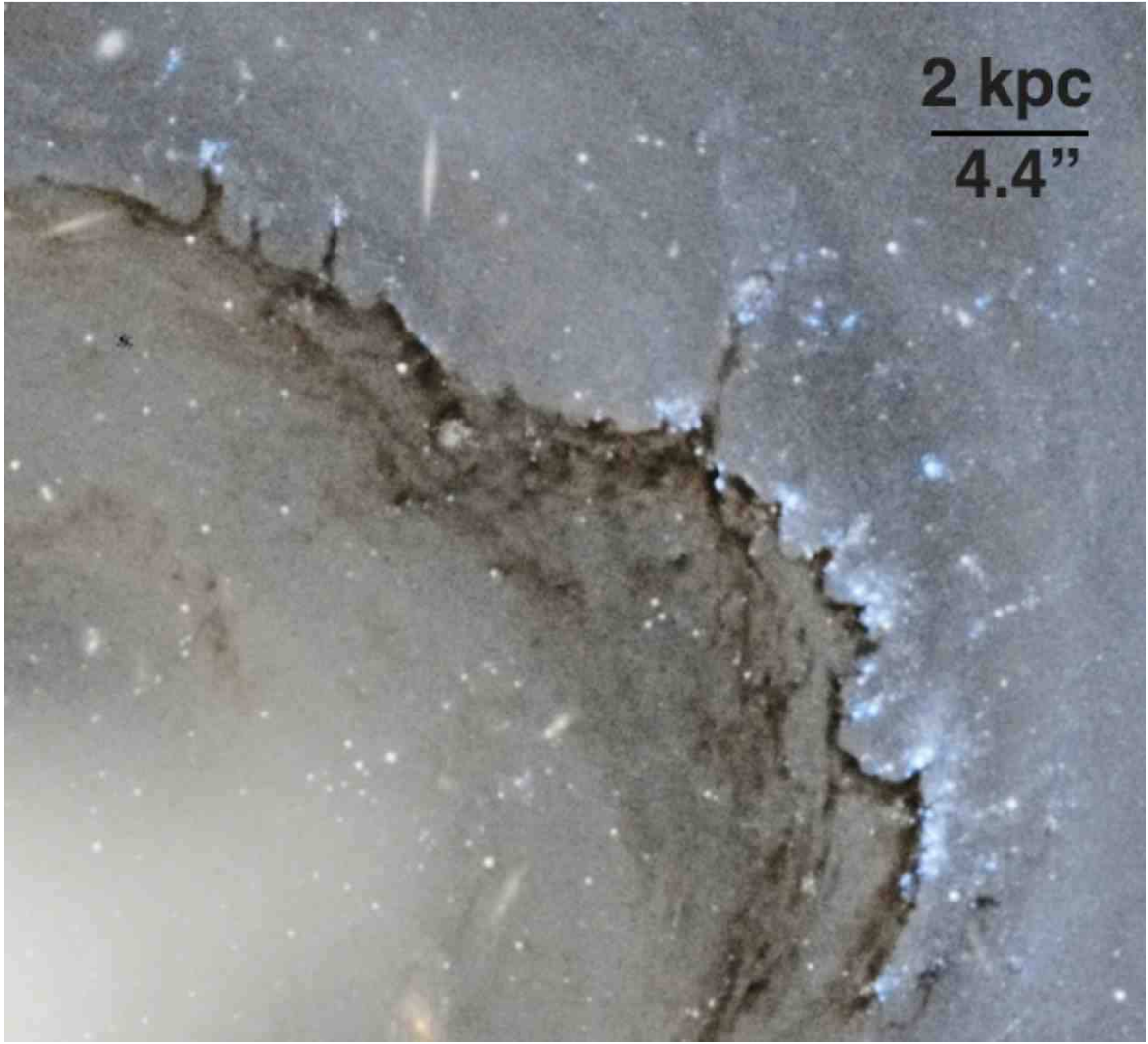}
\caption{
Leading (NW) side of the disk, showing the effects of strong ram pressure. The leading side harbors a 20-kpc long continuous dust front formed through ram pressure. The upper left (NNW) shows 3 linear head-tail filaments connected to a smooth dust front. The lower right (WNW) shows a 'dimpled curtain' of dust in a region with a high density of young stars. The continuity and substructure of this front, with dust filaments draped downstream from young star clusters, strongly suggests dense cloud decoupling partly inhibited by something which binds the ISM, possibly magnetic fields.
\label{fig5}
}
\end{figure*}

\begin{figure*}[ht]
\plotone{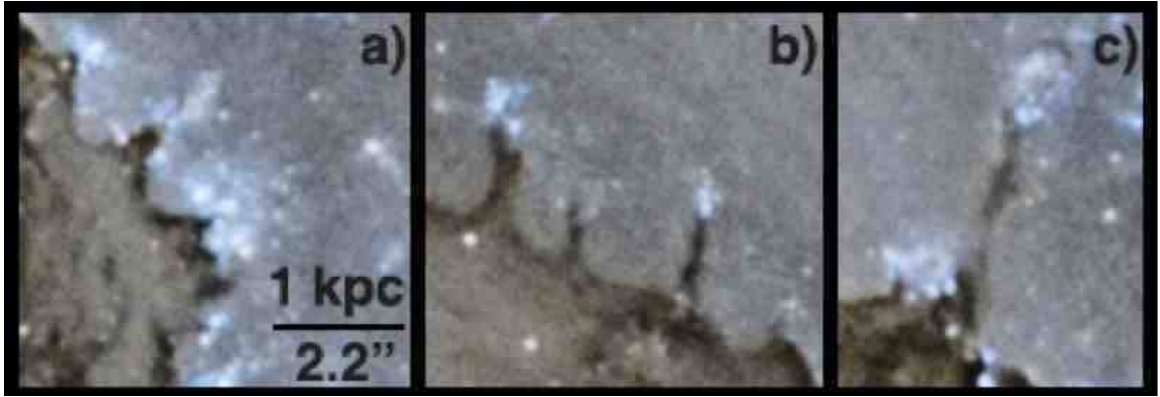}
\caption{
V-shaped, Y-shaped, and nearly linear head-tail filamentary features along the long leading side dust front. We interpret these as dense clouds which have partly decoupled from the surrounding ISM, but remain connected it by magnetic fields. The 3 panels from left to right suggest an evolutionary sequence of increasing decoupling. The region in the first panel, which has a ``dimpled curtain'' morphology, may also have higher gas density.
\label{fig6}
}
\end{figure*}

\begin{figure*}[ht]
\plotone{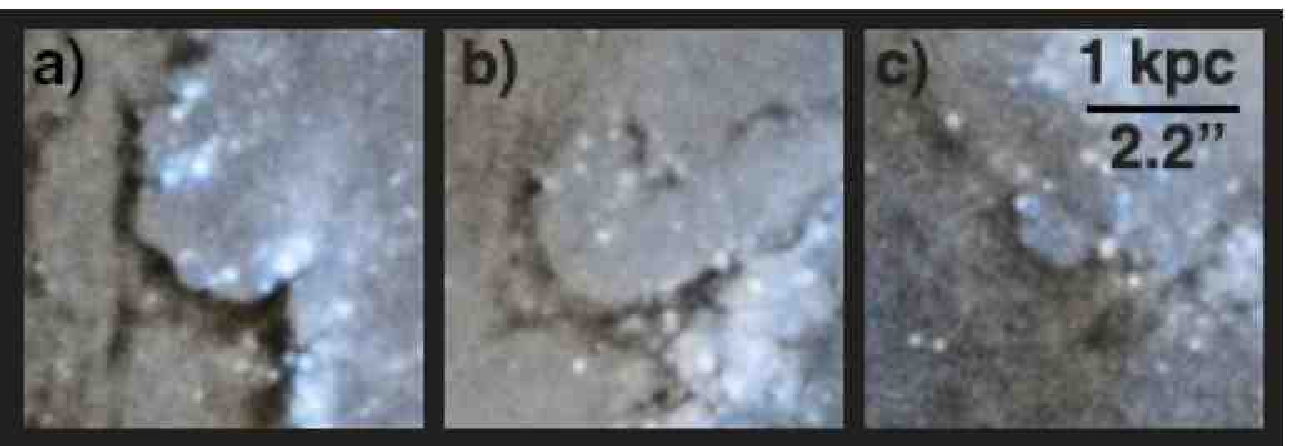}
\caption{
C-shaped filaments which divide dusty from largely dust-free regions.  Those in panels a and b are along the long leading side dust front, whereas the one in panel c is in the southern part of the galaxy. Note that the filaments are curved even though the ram pressure wind is straight. We interpret these as leading edges of ram pressure, where low density gas pushed by ram pressure remains connected by magnetic fields to decoupling higher density gas clumps. Note 2 tiny linear filaments (each with size $\sim$130x70 pc) emerging from young star complexes just upstream of C-shaped filament in panel a.
\label{fig7}
}
\end{figure*}

HST images (Figures 4-7) of NGC~4921 reveal a variety of remarkable dust extinction features that also strongly suggest that the galaxy is being actively stripped by ram pressure from the NW. Figure 4 shows dramatic azimuthal variations in dust morphology which we interpret as the combined effects of galaxy rotation and shielding of the trailing side. In the NW, where the HI distribution is most compressed and which is presumably the leading side of the disk, there is a well-defined, continuous-appearing front of dust that extends over 90 degrees and 20 kpc (Fig 5). Very little dust extinction is viewed upstream of this front. In contrast, the trailing side (SE) of the disk (Fig 4) shows only modest influences from ram pressure, presumably due to the effects of shielding from intervening ISM gas between the trailing and leading sides. 
The outer disk beyond the dust and HI is relatively blue, indicating that star formation occured there until recently, and that ram pressure stripping is quenching star formation from the outside in.
In this paper we focus on the leading (NW) side, which shows the strongest effects of ram pressure, including evidence for dense cloud partial decoupling and magnetic binding of high and low density gas. Other regions and topics will be discussed in a separate paper.

Dust extinction features generally correspond to the cool-cold ISM, or regions of atomic and molecular gas. We have estimated the extinction in selected regions by comparing the surface brightness in dusty and nearby dust-free regions. Extinction values in the F606W band range from $\leq$0.05 to 0.6 magnitudes, when averaged over regions of at least 10 pixels.  Assuming dust properties and a dust-to-gas ratio like that in the Milky Way, and a uniform layer of dust in front of all the stars, this translates to a minimum gas (HI+H$_2$) surface density of 1-10 M$\solar$ pc$^{-2}$. The true gas and dust surface density is undoubtedly much higher than this in some regions, since stars and dust are mixed together, and the dust distribution is highly inhomogenous within each pixel. \citet{ak14} estimate a likely correction factor of 10 for small decoupled dust extinction clouds in HST images of two Virgo cluster spirals. Thus densities in regions with visible dust extinction in the HST image of NGC 4921 probably range from 1-100 M$\solar$ pc$^{-2}$, and the darkest dust features are likely molecular clouds.

We interpret the long dust front in Fig 5 as ISM swept up by ram pressure along the leading side of an ICM-ISM interaction. The dust front appears contiguous over 20 kpc but it is hard to know the 3D structure. The large apparent difference between the leading and trailing sides in NGC~4921 indicates that the disk-ICM wind angle has a significant component in the disk plane. This component is more effective in pushing the disk ISM together into a swept-up front. Since NGC~4921 has a line-of-sight velocity which is blueshifted by 1500 km/s with respect to the cluster (NED), there must also be a significant wind component perpendicular to the disk, and the downstream direction is angled out of the disk to the far side.

The long leading side dust front is related to at least 2 spiral arms. The N part of the dust front is located near a continuation of a spiral arm which extends from the NE, but the W part of the dust front is located near a different spiral arm segment which is located further out in radius. In the NW, the dust front crosses from one spiral segment to the other, making a large change in galactocentric radius. So it is not simply spiral structure which is making this continuous ISM feature, but rather ram pressure plus possibly magnetic binding (discussed further below), acting on an ISM with pre-existing spiral arms. It makes sense that much of the ram pressure dust front would be near spiral arms, since this is where the ISM gas is densest and most resistant to ram pressure acceleration.

\section{Substructure in Leading Side Dust Front}

\subsection{Dense Cloud Decoupling}

The long leading side dust front shows significant substructure 
correlated with the density of young stars. 
In the W part of this dust front where there is a high concentration of young star clusters, the dust forms a nearly continuous ``dimpled curtain'', with a connected series of curved filaments (Fig 6a) extending downstream from knots of young stars. With respect to the knots, the dust forms downstream filaments with a 
curved V or Y-shape. Between adjacent knots, the curved filament generally forms a C-shape (Fig 7). Both are part of the continuous dust front, which appears as a connected multi-head structure.

Away from young star clusters the dust front is often smooth and straight, suggesting that in such regions virtually all the ISM has been swept up. 
In the NNW part of the dust front, which is a region with only a few star clusters, 
the filament morphology is different from the dimpled curtain region. Protruding outwards nearly perpendicularly from an otherwise smooth dust front are 3 nearly parallel dust filaments with lengths of 0.5-1 kpc and widths of 100-200 pc (Fig 6b). The 2 largest filaments have head-tail morphologies with complexes of young stars at the heads.  While the upper parts of the filaments are nearly linear, they appear to physically connect to the dust front. This is especially clear in the northernmost filament where, at the base, the 2 sides of the filament curve outwards to become the dust front. This is evidence that the linear filaments are physically connected to the dust front.
\footnote{
The PAs of filaments should reflect the local wind direction, modified by differential rotation in the disk, whose effect should tilt the filaments counterclockwise (toward more positive PAs), since the galaxy rotates clockwise. The PAs of the 4 long filaments in the NW range from -2 to -17 deg, with an average value of -9 deg. This suggests a wind angle similar to the PA of maximum HI asymmetry of 335 deg.
}

We propose that this substructure reflects spatial variations in ram pressure acceleration, due principally to variations in gas density, leading to the partial decoupling of the densest clouds. 
According to the widely cited and simple \citet{gg72} ram pressure stripping formula, 
gas can be accelerated if the ram pressure P$_{\rm ram}$ = $\rho$ v$^2$ exceeds the gravitational binding force per unit area
$\Sigma$$_{\rm gas}$ $\nabla$$\phi$, where $\nabla$$\phi$ is the gradient of the gravitational potential $\phi$ in the wind direction. The acceleration depends directly on the gas surface density $\Sigma$$_{\rm gas}$. At a given location in a galaxy, low density gas should be accelerated much more than high density gas, in the absence of other forces. 
The upstream vertices or heads are the regions most resistant to acceleration, and given their frequent association with young star clusters, are presumably the densest gas concentrations. While stars could contribute significantly to the total mass of the decoupling cloud, the masses are probably generally dominated by gas, since the star formation efficiency of molecular clouds is low.  As the surrounding lower density gas is accelerated downstream by ram pressure, it begins to separate and decouple from the dense unaccelerated gas clouds. 
But something appears to link the high and low density parts of the ISM, inhibiting the decoupling, and we discuss this in the next section.

There are similarities between the ISM features in NGC~4921 and elephant trunk features in HII regions, although the linear scale differs by a factor of 1000 \citep{carl03,carl13}. Both are regions of higher gas density than their surroundings, so they offer greater resistance to disturbances. In the HII regions it is photoionization, photoevaporation and photoheating from nearby luminous stars that ionizes and pushes away the lower density gas. In the cluster galaxies it is ram pressure from the ICM wind that pushes away the lower density gas. Magnetic fields acting to bind the ISM might be important for both, as suggested by Carlqvist \citep{carl03,carl13}.

\subsection {Evidence for Magnetic Binding in the ISM}

\begin{figure}[ht]
\plotone{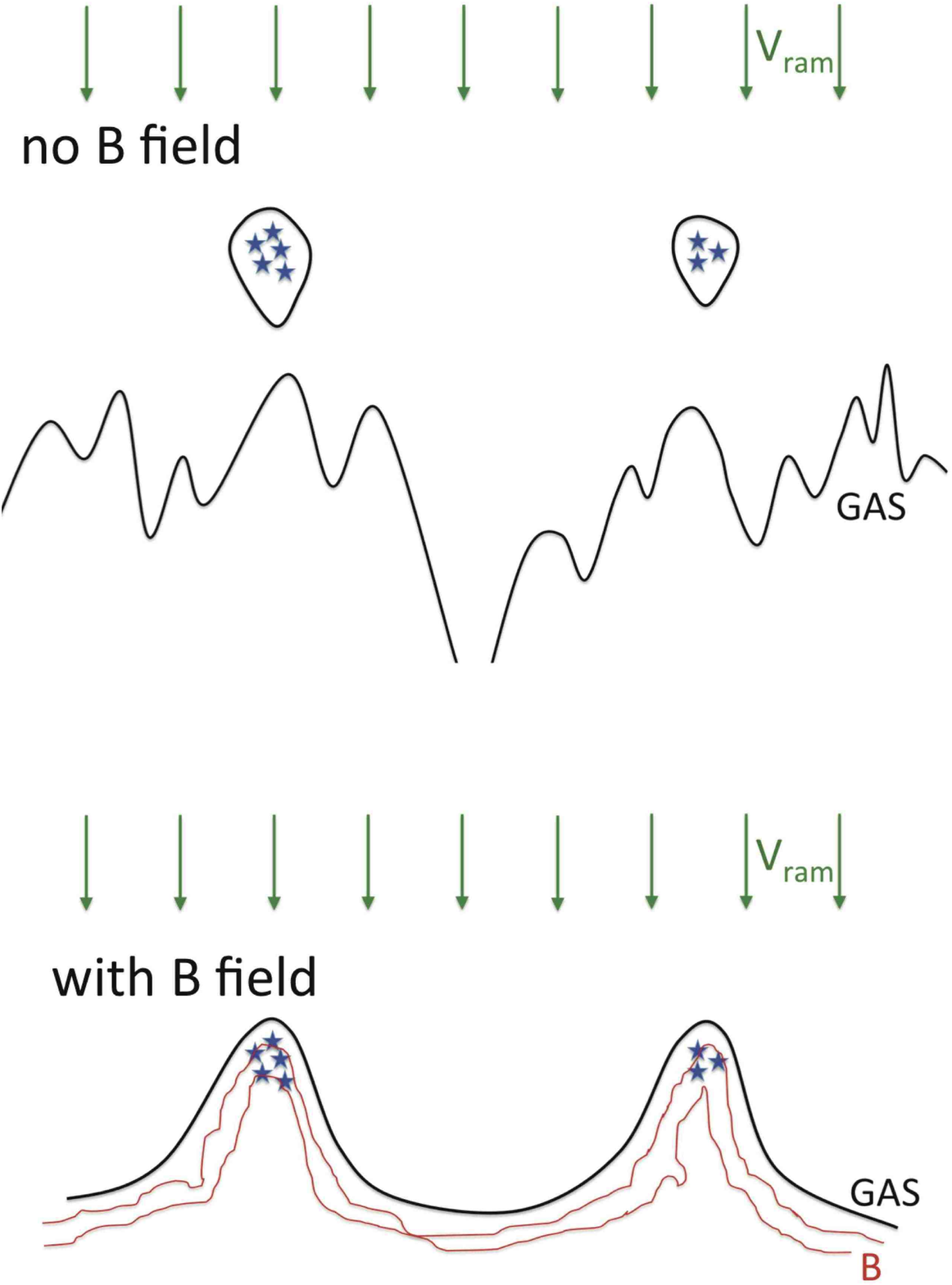}
\caption{
Cartoon illustrating proposed magnetic binding effect.
In the top panel there is no magnetic field, and the ram pressure wind accelerates gas parcels according to their surface density, which varies irregularly with position. The ISM front is irregular and dense clouds can fully decouple.
In the lower panel there is a magnetic field, which binds nearby high and low density gas parcels together. When acted upon by ram pressure, the acceleration of a gas parcel also depends on the magnetic binding to nearby gas parcels, smoothing out the ISM front and making it continuous. Dense clouds too dense to be accelerated remain connected to surrounding lower density gas by the magnetic field.
\label{fig8}
}
\end{figure}

\begin{figure}[ht]
\epsscale{0.87}
\plotone{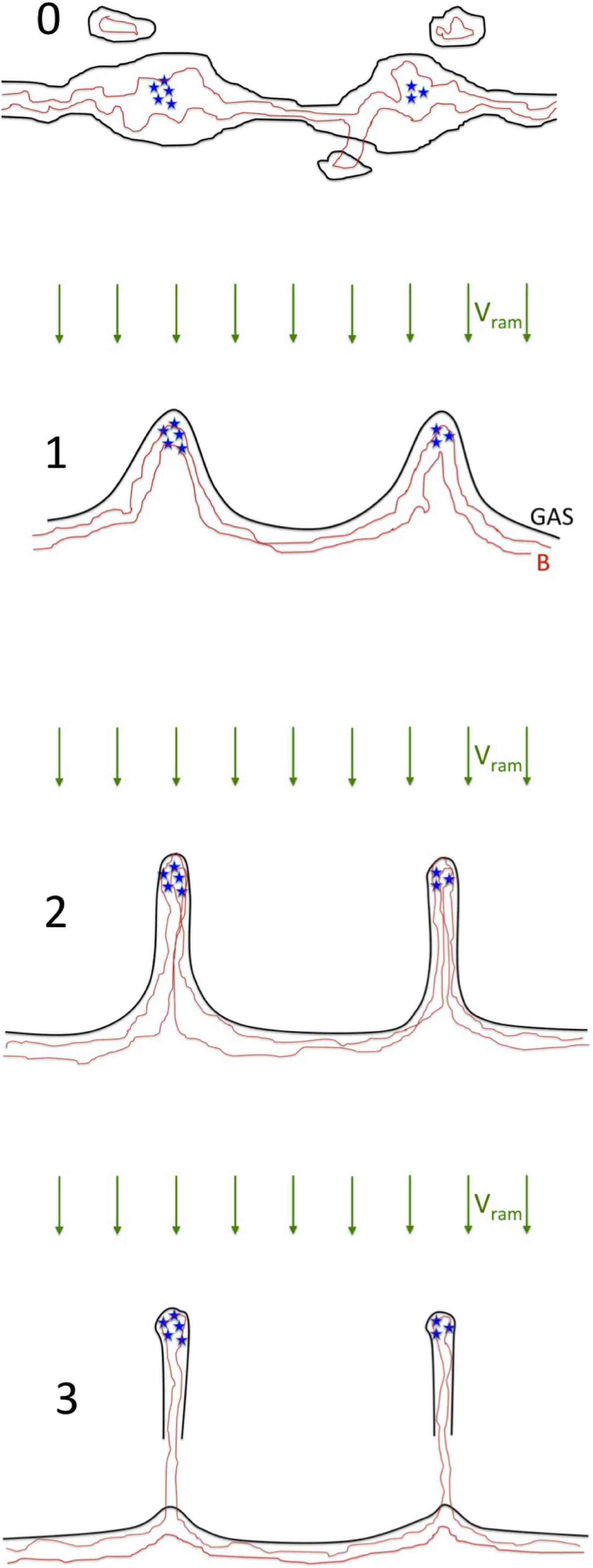}
\caption{
Cartoon illustrating proposed evolutionary sequence of cloud decoupling with magnetic binding.
0. Pre-ram pressure ISM permeated by magnetic fields and with dense clouds hosting star formation.
1. Early stage in dense cloud decoupling during ram pressure. Low density ISM is accelerated downstream but dense clouds are not so they start to decouple. However magnetic binding links the high and low density gas resulting in a continuous and regular ISM front. In this early stage, the gas and dust filaments curve away from the dense clouds, forming curved V-shaped features when viewed from the dense clouds, and C-shaped features when viewed in between the dense clouds.
2. Intermediate stage in dense cloud decoupling during ram pressure. The ISM front is accelerated further downstream, and the filaments connecting the decoupling dense cloud to the downstream front stretch and become straight near the head but remain curved near the ISM front, forming a curved Y-shaped filament (left). 
After further stretching the head-tail filament becomes straighter and nearly linear (right).
3. Later stage in dense cloud decoupling during ram pressure. Dense cloud in process of decoupling from receding ISM front. Filament has become so stretched out that field lines begin to disconnect. The densest parts of the filament should survive the longest.
\label{fig9}
}
\end{figure}

The morphology of the dust lanes affected by ram pressure, in particular the smoothly curved C-shaped features anchored near young star complexes, the nearly linear filaments connected to the perpendicular dust front, and the relatively smooth and straight lanes away from young star complexes, strongly suggests that magnetic fields link and partly bind distant parts of the ISM, and are dynamically important during ram pressure stripping.

At a given location in a galaxy, low density gas should be accelerated by ram pressure much more than high density gas, in the absence of other forces. Since the gas surface density varies irregularly with position, this would predict highly irregular stripping fronts, as shown in the cartoon in the upper panel of Fig 8. But this is not what is observed. Many dust fronts in NGC~4921 are smoothly curved, including the C-shaped features between adjacent vertices, indicating that the gas acceleration at a given location depends on neighboring positions. This is precisely the behavior expected if magnetic fields link adjoining gas parcels, as illustrated in the lower panel of Fig 8.

The C-shaped filaments provide evidence for magnetic binding. Kiloparsec-scale C-shaped filaments are in the long dust front as well as elsewhere in the galaxy, and all divide dusty from largely dust-free regions. There are no full circular shells, only partial shells. There are no large concentrations of young stars in the center of these C-shaped filaments, as would be expected if outward radial forces from young stars, evolved stars, or supernovae were making interstellar bubbles. Rather there are knots of young stars at the 2 ends of most of the C-shaped filaments. The symmetry axes of all these C-shaped filaments are the same within 20 degrees, and point towards the NW. Fig 7c shows a C-shaped filament from a region $\sim$15 kpc S of the nucleus, closer to the trailing side of the disk. It has a symmetry axis similar to those on the leading side, suggesting a common wind direction from the NW. Note that the filaments have curved shapes even though the ram pressure wind is straight.  While hydrodynamical instabilities acting at the boundary of 2 uniform media can create curved features, the fact that the C-shaped filaments are anchored by young star complexes suggests that they result from ram pressure acting on ISM with pre-existing substructure, rather than from instabilities. Indeed, the portions of the dust front in the NNW lacking young star clusters, and therefore with less substructure, are relatively straight.

The kpc-scale head-tail filaments in the NNW of NGC~4921 are similar to those previously seen in the ram pressure stripped Virgo cluster spiral NGC~4402. \citet{crowl05} interpreted these as dense clouds which decoupled from the surrounding lower density gas that experienced greater acceleration from ram pressure, which we think is still correct. However they also proposed that the elongated filament was formed from hydrodynamical ablation from the edges of the dense cloud. The dust morphology of NGC~4921 strongly suggests that it is not ablation or shadowing that forms the "tail" of these elongated filaments, but magnetic binding of high and low density regions of the ISM as ram pressure selectively accelerates lower density gas.

We interpret long smooth dust fronts as regions in which there are not many concentrations of sufficiently high density to decouple from the stripped downstream flow. The existence of long smooth dust fronts is suggestive of magnetic fields, as theoretical expectations and simulations show that strong fields suppress fragmentation and mixing by stabilizing the interface at the cloud surface \citep{frag05, shin08, yama11}.

\subsection{Evolutionary Effects on Substructure}

Along the dust front dense clouds are seen in various stages of inhibited decoupling. We suggest that the three regions highlighted in the panels of Figure 6 outline a possible evolutionary sequence. In Fig 9 we show a cartoon of the different evolutionary stages in our proposed model of dense cloud decoupling.

Dense decoupling clouds form the heads of V-shaped, Y-shaped, and linear filaments protruding upstream from a front of gas and dust.  In the early stages of decoupling, the 2D projected shape of the 3D dense gas distribution extending from a decoupling cloud has a curved V-shape (Fig 6a). As the front moves further downstream, it becomes more Y-shaped, with greater curvature near the base than the head (left filament in Fig 6b). At later stages, the part near the head becomes lengthened and nearly linear, due to greater stretching (right and center filament in Fig 6b). 
The Y-shape need not involve twisting of the magnetic field lines, as proposed by Carlqvist (2013), although that may well occur in some instances. A Y-shape can be caused by greater stretching of the field lines near the head than the base, and this will occur naturally during the decoupling.

If we look instead at the morphology of the 2D projected dust distribution in between 2 adjacent decoupling clouds, we find it is often C-shaped, where the C is formed from the bases of the curved V or Y filaments extending from adjacent dense clouds. This shape is common in regions with a high density of star formation (and at early stages of decoupling). In regions with a low density of star formation (or, at later stages of decoupling), the dust front has less curvature and in same places is nearly straight. 

The dimpled curtain region in the WNW (Fig 6a) is at an early stage, with dust filaments curving away from the heads with a large opening angle. The three linear filaments in the NNW (Fig 6b) are at a later stage, with the surrounding ISM pushed further downstream. The dust filaments emerging from the heads are nearly straight, with a small opening angle.

The longest head-tail filament, shown in Figure 6c, is $\sim$2 kpc in extent and is likely at a later evolutionary stage of decoupling than the 3 linear filaments in Fig 6b. 
The filament differs in morphology from the 3 northern ones in Figure 6b in that the dust density is very low $\sim$1/3 of the way between the dust front and the head. This suggests a later stage of decoupling, with a gap forming between the decoupled dense cloud and the dust front, and some gas near the base of the filament getting pulled into the dust front.  
We suspect that at later evolutionary stages most of the dense concentrations within each large filament ultimately become stranded between the decoupled head cloud and the receding dust front, forming a linear string of small dense clouds, or a "fossil filament". A possible fossil filament is observed in the Virgo galaxy NGC~4402 \citep{ak15}.

\subsection{Density Effects on Substructure}

\begin{figure*}[ht]
\plotone{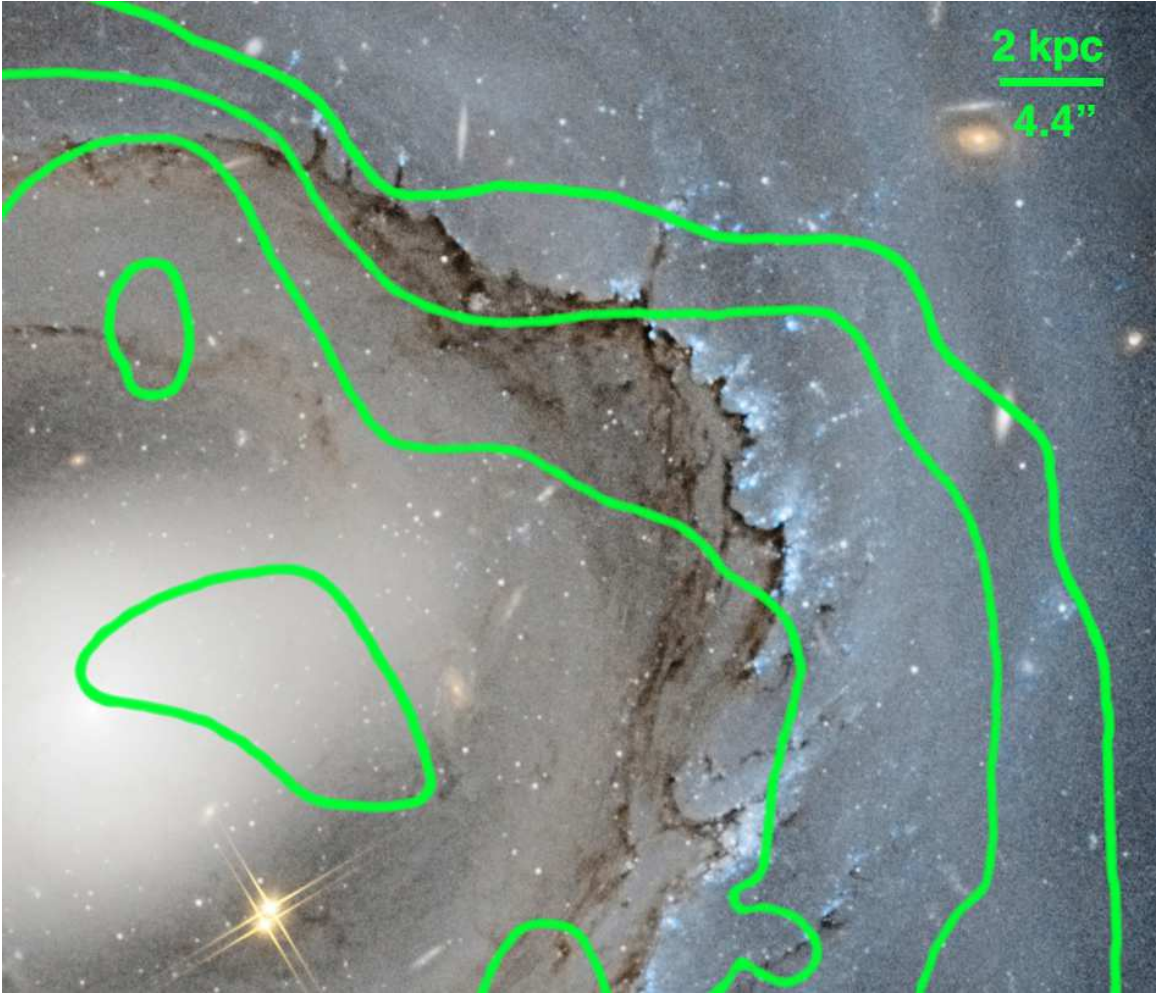}
\caption{
HI intensity contour map on color HST image (F606W+F814W) of NGC~4921, zoom-in on NW region featuring leading side dust front. Contours as in Fig 1. Note the relative lack of HI and young stars in the NNW region of the linear filaments and the relative abundance of HI and young stars in the NW region of the dimpled curtain.
\label{fig10}
}
\end{figure*}

If the only variable were the time exposed to a strong ICM wind, then the features in the W part of the dust front would be at a later evolutionary stage than those in the N. This is because the ISM in the W has been exposed to the full ICM wind for a longer time than the ISM in the N, assuming that spiral arms trail so the galaxy rotates clockwise. The different evolutionary stages of dense cloud decoupling proposed above do not vary smoothly with azimuthal angle along the dust front, as would be expected if the time exposed to the ICM wind was the only variable along the dust front. 

We think pre-existing substructure also varies enormously along the front, and is responsible for many of the differences along it. In the HST images, the main observable difference between the linear filament region (NNW) and the dimpled curtain region (WNW) of the dust front is the density of young star complexes, which presumably traces the the average gas density and the density of high density clouds. The linear filaments are in a region with isolated star-forming clouds and low average gas density, whereas the dimpled curtain is in a region with a high concentration of star-forming clouds and a high average gas density. In a 10 kpc length annular segment centered on the heads of the 3 linear NNW filaments, there are only ~3 star forming complexes (including the complexes at the heads of 2 of 3 the linear filaments), however over a similar-sized annular region centered on the dimpled curtain, there is $\sim$50 times more optical luminosity from young star complexes.

Support for a difference in the gas content of the 2 regions comes from an HI-HST overlay in the NW quadrant, shown in Figure 10. In the NE and W, the HI contours extend a bit beyond the main dust annulus, indicating that the HI itself extends to about the main dust annulus, and there is very little HI beyond. (Because of the low resolution of the HI data, the outer contours can be beyond the actual HI location.) HI has the shortest extent in the N to NNW, where the outermost HI contour barely reaches the leading dust front, implying that there is not much HI associated with this part of the long dust front. (If there were a significant concentration of HI on this part of the dust front, the HI contours would extend $\sim$15$''$ beyond the dust front.) This part of the dust front contains the 3 kpc-scale linear filaments. The HI appears least extended at the rightmost of the three filaments. The large difference in HI content of the 3 filament (NNW) region compared to the dimpled curtain region (WNW) of the leading dust front is consistent with a large difference in the amount of pre-existing dense gas in the 2 regions.

We note several systematic differences in the properties of the two regions. Relative to the dimpled curtain region, the linear filament region has a greater separation between adjacent heads, a larger distance between the decoupled heads and the most downstream part of the dust front, a filament shape which is less curved downstream from the heads, a lower dust density in the dust front, and a smoother dust front. All of these differences could be a consequence of the pre-stripping gas density on kpc scales. This is evidence that pre-existing substructure greatly affects the stripping process.

Just S of the dimpled curtain the dust front forms a big C-shaped inward filament (Fig 7a) in which dust has been pushed further downstream. This occurs just where the density of young stars drops, so probably indicates that a relative lack of dense gas has allowed ram pressure to push the gas and dust further downstream. In general along the dust front, dust is pushed further downstream in regions with fewer young stars, and this is also seen in the region between the largest head-tail filament and the group of 3 shorter ones.

\section {Simulations of Cloud-Wind Interactions and Evidence for the Importance of Magnetic Fields }

There have been numerous recent numerical simulations of clouds interacting with winds and shocks under various conditions \citep{frag05, mel06, shin08, coop09, pitt09, leao09, vanloo10, pitt11, jz13, aluz14}, but most do not produce the type of features that we observe.

Many simulations form an elongated tail behind a dense head \citep{shin08, pitt11, jz13}, and some produce multiple elongated filamentary features \citep{coop09}, but none of these connect to a perpendicular ISM front. The effects of a wind interacting with a fractal (multi-density) cloud without magnetic fields has been explored in simulations by \citet{coop09}, who find that the gas density distribution exhibits a highly irregular front after the onset of the wind, similar to the cartoon with no B field in the upper panel of Figure 8. Several simulations have explored the effects of magnetic fields. Even weak magnetic fields can drastically alter the evolution of the cloud compared to the hydrodynamic case \citep{shin08, vanloo10}. Strong fields suppress instabilities, and therefore suppress fragmentation and mixing by stabilizing the interface at the cloud surface \citep{frag05, shin08, yama11}. Radiative cooling has a large impact on cloud evolution \citep{coop09, tb10}, and there is significant interplay between magnetic fields and radiative cooling \citep{frag05, jz13}. Recent simulations of shocks interacting with multiple magnetized clouds  \citep{aluz14} show that behavior is complex and depends upon the initial positions of the clouds and the orientation and strength of the magnetic field.

The only cloud-wind interactions we are aware of that show a head-tail morphology with the tail connected to a downstream perpendicular gas front are those of \citet{leao09}, in which a strong wind (supernova shock) encounters a magnetized uniform cloud embedded within a lower density magnetic gaseous medium. Figures 4 and 6 of \citet{leao09} show morphologies resembling the linear and Y-shaped features in NGC~4921.  Similar simulations but without magnetic fields by the same group \citep{mel06} do not show such morphologies. We consider this as evidence that magnetic fields are important to the dust morphologies in NGC~4921. Details such as the diameter ratio of the head and tail, and the length to diameter ratio of the tail do not match, but these probably depend on various parameters which do not match in the simulations and the galaxy, such as shock strength and ISM sub-structure. C-shaped features are not seen in the \citet{leao09} simulations, but these require 2 adjacent dense clouds (heads), which are not modeled.

We think the V,Y, linear and C-shaped filamentary substructure along the leading edge dust front arise from partly decoupled, magnetically linked clouds that form from a strong wind interacting with a magnetized, multi-density ISM. We think the interaction can be characterized by a density contrast within the ISM of at least 10$^2$, a density contrast between the wind and the lowest density ISM of $\sim$10$^3$, a ram pressure sufficient to move the low density but not the high density ISM, and a magnetic energy density in compressed regions comparable to the ram pressure.

\section {Discussion}

In order for a magnetic field to be dynamically important, the magnetic energy density must be comparable to the ram pressure. Adopting n$_{\rm ICM}$ = 10$^{-3}$ cm$^{-2}$ \citep{briel92} and v = 2000 km/s, the ram pressure is P$_{\rm ram}$ = $\rho$v$^2$ = 7x10$^{-11}$ dyn cm$^{-2}$, and the magnetic energy density B$^2$/8$\pi$ is equal to the ram pressure for a magnetic field strength of 40 $\mu$G. In nearby grand-design spiral galaxies with massive star formation, B=15$\mu$G is a typical average strength of the total field, whereas in the density-wave spiral arms of M51 the total field strength is 25$\mu$G \citep{bw13}. Compression of the gas by ram pressure is expected to increase the field strength above that seen in a typical spiral arm, so a magnetic field strength of 40$\mu$G at the leading side of NGC~4921 seems reasonable.

If this interpretation of the ISM substructure of NGC~4921 is correct, future observations should show that the magnetic energy density is comparable to the ram pressure, and that the magnetic field is aligned along the leading edge dust front. Observations of many Virgo spirals experiencing rps have shown ridges of enhanced linear radio polarization along the leading edges \citep{voll07, voll08b, voll13a}, with magnetic fields aligned with the leading edges. These polarized ridges are thought to be where the gas and magnetic fields in the disk and halo of the galaxy get compressed by ram pressure, increasing the degree of large-scale polarization. Further evidence that ISM compression produces the polarized ridges comes from radio deficit regions \citep{mur09} located just outside the polarized ridges. Radio deficit regions are thought to be where the magnetorelativistic plasma on the leading side is pushed into the galaxy.

This magnetic binding of nearby gas parcels means that low surface density gas parcels should be harder to ram pressure strip than the simple Gunn \& Gott formula predicts, and higher density parcels are easier to ram pressure strip.  The Gunn \& Gott formula probably works better if averaged over an appropriately large area that includes high and low density regions. There is obviously a limit to the magnetic linking, since the observations show that denser gas concentrations can ultimately decouple from the rest of the ISM. The magnetic binding of low and high surface density regions of the ISM should make it more difficult for ram pressure to punch holes through low surface density regions of the disk, as seen in some simulations without magnetic fields \citep{qmb00}. Punching through holes in the disk can increase the efficiency of hydrodynamical stripping by increasing the surface area exposed to instabilities. Thus magnetic binding could suppress hole-punching and therefore decrease the efficiency of hydrodynamical stripping.

The stripping susceptibility of higher density clouds is affected in 2 different ways by dynamically important magnetic fields. Magnetic coupling can make it easier to directly strip dense clouds by ram pressure. But it makes it harder to ablate the decoupled clouds by hydrodynamical processes since magnetic fields should inhibit the KH and RT instabilities which can otherwise destroy dense star-forming clouds \citep{frag05, shin08, yama11}. 

While several simulations of ram pressure stripping on galaxy scales have included magnetic fields in various ways \citep{ov03, rus14, ts14}, none have yet been done at the spatial resolution required ($\leq$50 pc) to see the decoupled dense clouds. \citet{ts14} concluded that magnetic fields in the disk did not greatly change the overall stripping rate with a face-on wind
but the 150 pc resolution of the simulations is insufficient for seeing the substructure on the scales revealed in the HST images of NGC 4921. It is important to resolve the dense gas concentrations in the ISM, since these are most resistant to stripping and can act as magnetic anchors on surrounding low density ISM. Simulations compared with observations remain the best way to estimate the efficiency of stripping, and a successful simulation should reproduce the ISM morphology seen in strongly stripped galaxies like NGC~4921.

\section {Summary}

NGC~4921 is the most massive spiral galaxy in the Coma cluster, located at 35\% of the cluster virial radius.
Dust extinction features in deep HST images show evidence of strong ongoing ram pressure stripping.
Since the galaxy is viewed nearly face-on, it shows the behavior of the disk ISM as it is stripped
more clearly than previously known stripped galaxies. NGC~4921 contains only 10\% the HI of a normal spiral.  VLA mapping shows the HI disk is truncated well within the stellar disk, and is highly asymmetric, with an extent twice as large in the SE, suggesting that ram pressure is acting from the NW.  The galaxy is presumably moving toward PA$\simeq$335$\deg$, an offset of 55$\deg$ from the direction toward cluster center (at PA=280$\deg$), implying that the galaxy is approaching the cluster center on its orbit, and ram pressure is increasing. HI kinematics show non-circular motions in the outer northern part of the galaxy, presumably due to ram pressure.

Where the HI distribution is truncated and compressed in the NW, the HST image shows a 20 kpc length dust extinction front, which we interpret as ISM swept up by ram pressure. There is little or no dust extinction beyond this dust front. The substructure in this dust front reveals important new details on what happens in the ISM during stripping. While parts of the long dust front are relatively smooth, in parts where there are young stars dust features (100 pc-1 kpc scale) protrude from the large scale dust front in the upstream direction. The heads of many but not all of these protruding features are coincident with associations of bright blue stars. We interpret the heads of these head-tail features as dense clouds that are too dense to accelerate by ram pressure, which decouple from lower density surrounding gas which gets accelerated downstream by ram pressure. The shape of the head-tail filaments and their apparent connection to the downstream ISM front strongly suggest that the high and low density gas remain partly bound during the stripping process, and that the decoupling is inhibited by something that binds the ISM together, which we propose is magnetic fields. We propose that as decoupling proceeds, the dense decoupled cloud forms the head of a head-tail feature. The tail first forms a curved V-shaped filament, then a curved Y-shaped filament, then a nearly linear filament, with the filament connecting to the downstream ISM front. In between adjacent dense decoupling clouds, the dust often forms a C-shaped filament, with the 2 dense clouds anchoring the 2 ends of the filament.

The morphology of the ISM substructure appears to depend on initial gas density as well as evolutionary stage. In the part of the long dust front that has less HI and fewer young stars, the dust front is generally smoother, and the head-tail filaments are more linear. In the part of the long dust front that has more HI and young stars, the dust front exhibits more substructure, and the head-tail filaments are more curved. We suggest this reflects the importance of pre-existing ISM substructure such as spiral arms in rps. 
The long leading side dust front appears to be composed partly from 
2 distinct spiral arm segments plus a connecting segment which crosses between the arms.
The part of the dust front with more HI and young stars is associated with one of the spiral arm segments.
We can't tell whether star formation is triggered or merely revealed along the leading side, although we don't see star formation in the compressed downstream dust front or other evidence for triggering.

The HST image suggests that magnetic fields are dynamically important during ram pressure stripping.
Evidence for magnetic binding comes from simulations of cloud-wind interactions on small ($<$10 pc) scales. The only simulations that produce linear head-tail features connected to a downstream perpendicular ISM front are ones with magnetic fields \citep{leao09}. Some recent simulations of ram pressure stripping in galaxies on large scales ($>$100 pc) have included magnetic fields. Based on their simulations, \citet{ts14} have concluded that magnetic fields do not greatly affect the efficiency of stripping. However the resolution in the galaxy simulations does not yet reach the scales where cloud decoupling and magnetic binding are observed in NGC~4921, thus the effects of ISM substructure and magnetic fields on the efficiency of stripping remain unsettled.


\acknowledgements

We are grateful to Roberto Colombari for permission to use his beautiful HST image of NGC~4921. We thank Pavel Jachym, Richard Larson, Stephanie Tonnesen, Xavier Koenig, and Jonathan Foster for helpful discussions and comments on the manuscript. We also thank the referee Jay Gallagher for a helpful report that helped us improve the paper. HBA thanks the Mexican CONACyT for the support through a research project (169225). The data presented in this paper were obtained from the Mikulski Archive for Space Telescopes (MAST). STScI is operated by the Association of Universities for Research in Astronomy, Inc., under NASA contract NAS5-26555. Support for MAST for non-HST data is provided by the NASA Office of Space Science via grant NNX13AC07G and by other grants and contracts. This research has made use of the NASA/IPAC Extragalactic Database (NED) which is operated by the Jet Propulsion Laboratory, California Institute of Technology, under contract with the National Aeronautics and Space Administration.  

\medskip

\facility\
{\it Facilities: } \rm \facility{HST} \facility{VLA}

\end {document}